\pdfoutput=1

\documentclass[RNAAS]{aastex62}
\usepackage{graphicx,epsf,soul,xcolor,gensymb}

\begin{document}

\title{Six years of luminous X-ray emission from the strongly interacting type-Ib SN\,2014C captured by Chandra and NuSTAR}

\correspondingauthor{Daniel Brethauer}
\email{danielbrethauer2021@u.northwestern.edu}

\author{D. Brethauer}

\affiliation{Department of Physics and Astronomy and CIERA, Northwestern University, 2145 Sheridan Road, Evanston, IL 60208, USA}

\author[0000-0003-4768-7586]{R.~Margutti}

\affiliation{Department of Physics and Astronomy and CIERA, Northwestern University, 2145 Sheridan Road, Evanston, IL 60208, USA}

\author[0000-0000-0000-0000]{D.~Milisavljevic}

\affiliation{Department of Physics and Astronomy, Purdue University, 525 Northwestern Ave., West Lafayette, IN. 47907, USA}

\author[0000-0000-0000-0000]{M. Bietenholz}

\affiliation{Hartebeesthoek Radio Astronomy Observatory, PO Box 443, Krugersdorp, 1740, South Africa}
\affiliation{Department of Physics and Astronomy, York University, Toronto, M3J 1P3, Ontario, Canada}


\begin{abstract}
We present the first coordinated soft and hard 0.3-80 keV X-ray campaign of the extragalactic supernova SN\,2014C 
in the first $\sim$2307 d of its evolution.
SN\,2014C initially appeared to be an ordinary type Ib explosion but evolved into  
a strongly-interacting hydrogen-rich type IIn SN over $\sim1\,\rm{yr}$. We observed signatures of interaction with a dense medium across the X-ray spectrum,
which revealed the presence of a $\sim\,1-2\,\rm{M}_{\odot}$ shell of material at $\sim6\times10^{16}\,\rm{cm}$ from the progenitor. This finding challenges current understanding of hydrogen-poor core-collapse progenitor evolution. Potential scenarios to interpret these observations include (i) the ejection of the hydrogen envelope by the progenitor star in the
centuries prior to the explosion; (ii) interaction of the fast Wolf-Rayet (WR) star wind with the slow, dense wind of the Red Super Giant (RSG) phase, with an anomalously short WR phase.

\end{abstract}

\section{Introduction}

SN\,2014C evolved spectroscopically from a type Ib to a type IIn supernova (SN) as a result of the shock interaction with a  
H-rich circumstellar material (CSM) shell \citep{Milisavljevic15}. The presence of a dense H-rich medium in close proximity to a H-poor core-collapse SN has been inferred in a number of cases, including normal type Ib/Ic SNe \citep[e.g., 2001em, 2004dk,][]{Chugai06,Mauerhan18} and superluminous SNe \citep[e.g., iPTF13ehe, iPTF15esb, iPTF16bad,][]{Yan17}. 
This phenomenology might require enhanced mass-loss $\lesssim \,2000\,\rm{yrs}$ before explosion due to physical mechanisms that deviate from the traditional picture of line-driven winds in single massive stars (e.g. wave-driven mass-loss, envelope ejections driven by nuclear burning instabilities or binary interaction), or, alternatively, unusually short WR phases of evolution \citep[e.g.,][]{Smith14}.

Here we update the analysis of \cite{Margutti2017} and present new broad-band X-ray observations of SN\,2014C up to $t=2307\,\rm{d}$ post explosion, enabling  
unprecedented insight into time-dependent mass-loss mechanisms. We study these mechanisms by using the SN shockwave as a probe of the environment as described in \citet{Chevalier17}.

\section{Observations and Analysis}
We observed SN\,2014C with \emph{Chandra} and \emph{NuSTAR} 
at $t\,=\,396-2307\,\rm{d}$ past explosion (December 30th 2013) as part of our joint monitoring program. Each \emph{Chandra} observation was paired with a \emph{NuSTAR} observation taken within $\Delta\,t/t<10^{-3}$ to provide a complete view of the hard and soft X-ray emission. 
We reprocessed the \emph{Chandra} and \emph{NuSTAR} data following standard procedures within \texttt{CIAO} and \texttt{NuSTARDAS}. We find evidence for bright X-ray emission in each of our exposures both at soft and hard X-rays energies.

For \emph{Chandra} observations, we extracted a spectrum from a circular region of 1.5\arcsec\, centered at the location of SN\,2014C.
We followed a similar procedure for \emph{NuSTAR} exposures, but used a 1\arcmin\, region.
We performed a joint spectral fit of each \emph{CXO+NuSTAR} observation with an absorbed thermal bremsstrahlung  model (\texttt{tbabs*ztbabs*bremss}) within \texttt{XSPEC}. 
We accounted for contamination from unrelated sources inside the \emph{NuSTAR} PSF following a similar method as \cite{Margutti2017}. \emph{NuSTAR} observations are heavily dominated by background emission at energies $>40$ keV and we thus restricted our spectral analysis to the 3-40 keV range.
We adopted a Galactic neutral hydrogen column density $NH_{\rm MW}\,=\,6.14\,\times 10^{20}\,\rm{cm}^{-2}$ \citep{Kalberla05} in the direction of SN~2014C and a distance of 14.7 Mpc \citep{Freedman01}.

\begin{figure}
    \centering
    \includegraphics[width=0.97\textwidth]{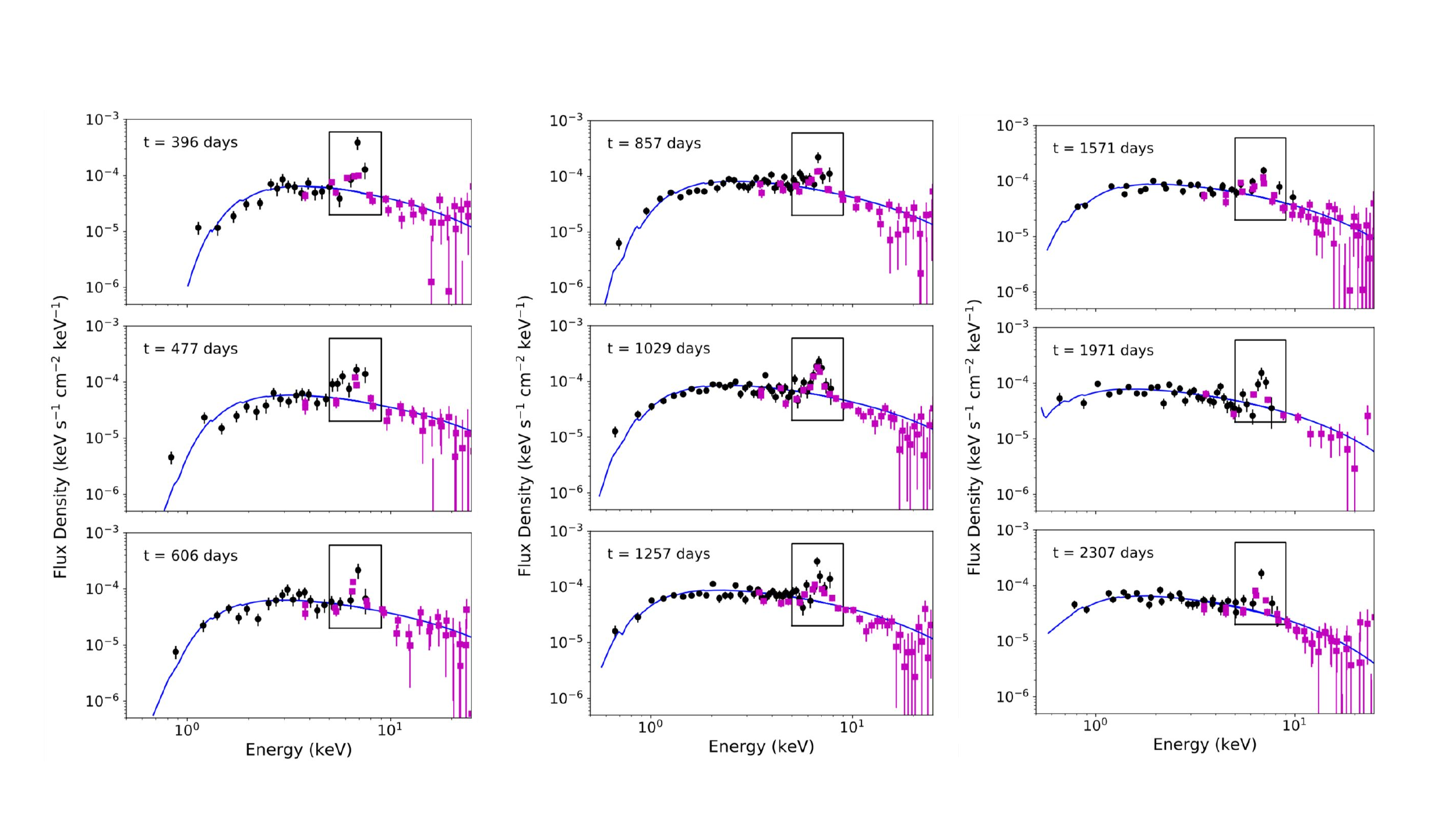}
    \caption{\emph{Chandra} (black circles) and \emph{NuSTAR} (magenta squares) broadband (0.5-40 keV) X-ray  spectra of SN\,2014C acquired at $t=396-2307\,\rm{d}$ post explosion. Thick blue line: best fitting absorbed bremsstrahlung model.
     Black rectangles: excess of emission with respect to this model  that we associate with K$\alpha$ line transitions in H-like or He-like Fe atoms. 
    }
    \label{fig:Xray}
\end{figure}

\section{Discussion and Conclusions}
A SN shock expanding into the CSM accelerates particles that cool down in the X-rays by radiating synchrotron and bremsstrahlung emission \citep{Chevalier17}. For shocks propagating in high-density media, like in SN\,2014C, thermal bremsstrahlung emission dominates over the synchrotron in X-rays. The observed thermal bremsstrahlung spectrum depends on the emission temperature $T$, (which directly  probes the shock velocity) and on the electron density of the radiating medium (and hence on the overall density $\rho$). Additionally, the presence of dense local neutral matter can lead to  photo-electric absorption that leaves a clear imprint at energies $\lesssim1\,\rm{keV}$  (Fig. \ref{fig:Xray}). We quantify
the local absorption with an intrinsic column density $NH_{\rm int}$. By fitting the thermal bremsstrahlung spectra of SN\,2014C over $\sim6\,\rm{yrs}$ of evolution, we constrained $T(t)$, $NH_{\rm{int}}(t)$, and the density profile encountered by the blastwave $\rho(r)$. 

We find $kT$ peaks at $\sim$23 keV at $ t\sim500 \,\rm{d}$, and cools as $T(t)\propto\,t^{-0.5}$. $NH_{\rm{int}}$ also declines with time as the shock propagates through the dense CSM shell. Specifically, we find $NH_{\rm{int}}(t)\propto\,t^{-1.4}$ starting from $\approx\,3\times 10^{22}\,\rm{cm^{-2}}$ at $t\sim\,400\,\rm{d}$. The resulting unabsorbed broad-band 0.3-100 keV X-ray light-curve peaks at $L_x\approx5.5\times 10^{40}\,\rm{erg\,s^{-1}}$ at $ t\sim1000\,\rm{d}$, followed by $L_x(t)\propto\,t^{-1}$, which is significantly less steep than what is expected from adiabatic expansion. Finally, we note the presence of an excess of emission at $\sim$6.7 keV (Fig.\ \ref{fig:Xray}) that we interpret as emission from H-like or He-like Fe atoms.

Combining the spectral parameters inferred above with direct measurements of the forward shock radius obtained from VLBI
observations of SN\,2014C by \cite{Bietenholz17} and Bietenholz et al., (\emph {submitted}), we constrain the properties of its environment. We find that
the progenitor star is surrounded by a low-density cavity (particle density $n<\,100\,\rm{cm}^{-3}$, \citealt{Margutti2017}) extending to $\sim\,6\,\times\,10^{16}\,\rm{cm}$. At $\sim\,6\,\times\,10^{16}\,\rm{cm}$, the blastwave encountered a dense shell of material with density $n\sim\,2\,\times\,10^{6}\,\rm{cm}^{-3}$ extending to at least $8\,\times\,10^{16}\,\rm{cm}$ and with a density profile beyond the shell $\rho(r)\,\propto\,r^{-2.5}$. Our observations sample the medium out to 
$r\sim2.5\times\,10^{17}\,\rm{cm}$.
Under the assumption of spherical symmetry supported by recent VLBI observations of Bietenholz et al, (\emph{submitted}) (see however \citealt{Milisavljevic15}), the total shell mass is $\sim\,1-2\,\rm{M}_\odot$ considering realistic filling factors.

Optical spectroscopy \citep{Milisavljevic15} indicates the CSM shell is H-rich, but it is unclear how this CSM shell was produced at such close distances from the explosion site. Viable scenarios fall under two broad categories. A first class of models does \emph{not} require erratic mass-loss. Rather,
the observed CSM shell is
the product of the interaction of a fast wind with a slow dense wind from the previous phase of stellar evolution, e.g. the transition from RSG to the WR phase (a known mechanism able to produce ``bubbles'' around Galactic WRs at typical distances $\sim$ tens of pc;  significantly larger than our inferred shell radius, e.g., \citealt{Marston97}). As a result, the phenomenology of SN\,2014C does require an anomalously short WR phase of $\lesssim$ 1700 yrs.  

A second
class of models invoke the ejection of the progenitor's H-rich envelope as a result of various mechanisms including 
binary interaction (\citealt{Margutti2017,Sun20} for the specific case of SN\,2014C), wave-driven mass-loss \citep{Quataert12,Shiode14,Fuller17,Fuller18,Wu20},  
or 
nuclear burning instabilities similar to those during O and Ne burning phases \citep{Arnett11,Shiode13,Smith14b}. 

Studies of statistical samples of SN\,2014C-like events across the electromagnetic spectrum hold the key to constraining the physics of the most spectacular mass-loss histories of evolved massive stars.

\acknowledgments

We are grateful to the entire \emph{CXO} and \emph{NuSTAR} teams for carrying out our observations, to the Office of Undergraduate Research at Northwestern University and the  
Illinois Space Grant for funding this research.

\bibliography{SN2014C}

\end{document}